\documentclass[11pt]{article} 

\usepackage{undertilde}
\usepackage[normalem]{ulem} 
\usepackage{amscd,amsmath}
\usepackage {amsfonts,apalike}
\usepackage{epsfig,graphicx,graphics}

\usepackage{cancel}
\DeclareGraphicsExtensions{.pdf,.png,.jpg}

\title{\vspace{-1.2in} Equivalent Theories and Changing {Hamiltonian} Observables  in General Relativity\footnote{Published in \emph{Foundations of Physics},  open access https://doi.org/10.1007/s10701-018-0148-1 }}

\author{J. Brian Pitts\\
 University of Cambridge } %

\date{30 June 2017}

\begin{document}

\maketitle 

\abstract{Change and local spatial variation are  missing in  Hamiltonian General Relativity according to the most common definition of observables as having $0$ Poisson bracket with all first-class constraints.  But other definitions of observables have been proposed. In pursuit of Hamiltonian-Lagrangian equivalence, Pons, Salisbury and Sundermeyer use the Anderson-Bergmann-Castellani gauge generator $G$, a tuned sum of first-class constraints.   Kucha\v{r} waived the $0$ Poisson bracket condition for the Hamiltonian constraint to achieve changing observables.  A systematic combination of the two reforms might  use the gauge generator but permit non-zero Lie derivative Poisson brackets for the external gauge symmetry of General Relativity.

Fortunately one  can test definitions of observables  by calculation using two formulations of a theory, one without gauge freedom and one with gauge freedom.  The formulations, being empirically equivalent, must have equivalent observables. For de Broglie-Proca non-gauge massive electromagnetism, all constraints are second-class, so everything is observable.  Demanding equivalent observables from gauge Stueckelberg-Utiyama electromagnetism, one finds that the usual definition fails while the Pons-Salisbury-Sundermeyer definition with $G$ succeeds.  This definition does not readily yield change in GR, however.

Should GR's  external gauge freedom of General Relativity share with internal gauge symmetries the  $ 0$ Poisson bracket (invariance), or is covariance (a transformation rule) sufficient?    A graviton mass breaks the gauge symmetry (general covariance), but it can be restored by parametrization with clock fields.  By requiring equivalent observables, one can test whether observables should have  $0$  or the Lie derivative as the Poisson bracket with the gauge generator $G$.  
The latter definition is vindicated by calculation.  While this conclusion has been reported previously, here the calculation is  given in some detail. }


\section{Problem of Missing Change and Spatial Variation in Observables}

Already in the mid-1950s there arose the problem of missing change in observables in the constrained Hamiltonian formulation of General Relativity:  ``There are indications that the Hamiltonian of the general theory
of relativity may vanish and that all the observables are constants of the motion.'' \cite{BergmannGoldberg} (see also \cite{AndersonChange}). 
This result appeared not too long after the introduction of novel distinctively Hamiltonian notions of gauge transformation and observables \cite[section 4]{BergmannSchiller}, in contrast to the previously manifestly Lagrangian equivalent work \cite{AndersonBergmann}.   
Besides the ``problem of time'' due to missing change \cite{IshamTime,KucharCanadian92}, which owes much to how the Hamiltonian constraint $\mathcal{H}_0$ is treated, there is also a  problem of space:  local spatial variation is excluded by the condition for observables  $\{O, \mathcal{H}_i\} =0$, pointing to global spatial integrals instead   \cite{TorreObservable}.

However, Bergmann  was prepared to define observables in a variety of inequivalent ways; while his definition in terms of first-class secondary constraints is intrinsically Hamiltonian, at times he wanted a definition that was independent of the Hamiltonian formalism   \cite{BergmannKomarRoyaumont,Bergmann,BergmannHandbuch}. 
Relatedly, Pons, Salisbury and Sundermeyer have proposed a reformed definition of observables using the Anderson-Bergmann-Castellani gauge generator $G$, a tuned sum of all first-class constraints including the primaries \cite{AndersonBergmann,CastellaniGaugeGenerator,PonsSalisburySundermeyerFolklore}.

At times Bergmann and others have expected  observables to be local or at least quasi-local  \cite[p. 250]{BergmannHandbuch} \cite[p. 115]{SmolinPresent}.  
Kucha\v{r}, in the interests of finding real change,  has been prepared to abolish altogether (not simply weaken) the requirement that observables have $0$ Poisson bracket with what generates time gauge transformations (which he took to be the Hamiltonian constraint $\mathcal{H}_0$)  \cite{KucharCanadian92,KucharCanonical93}.
According to Kiefer, 
\begin{quote} 
Functions $A(q,p)$ for which $\{A, \phi_A\} \approx 0$ holds are often called \emph{observables} because they do not change under a redundancy transformation.  It must be emphasized that there is no a priori relation of these observables to observables in an operational sense.  This notion was introduced by Bergmann in the hope that these quantities might play the role of the standard observables in quantum theory (Bergmann 1961).  \cite[p. 105; see also p. 143]{Kiefer3rd} \end{quote}

By implementing the inevitable  requirement that empirically equivalent theories have equivalent observables using the novel examples of  massive photons and (formally) massive gravitons, this paper and its predecessor \cite{ObservablesEquivalentCQG} reconsider the definition of observables and show that the conventional definition requires both the Pons-Salisbury-Sundermeyer reform to use $G$ rather than separate first-class constraints \emph{and} a novel non-zero Lie derivative Poisson bracket for external symmetries, partly inspired by Kucha\v{r}.  As a result, observables are local $4$-dimensional scalars, vectors, tensors, densities, \emph{etc}., just as in Lagrangian/geometric formulations, including the metric and the curvature tensors.  Thus change and local spatial variation are present after all.

The failure of observables to play their expected role has also led to circumvention by introducing new notions to do roughly the job that observables disappointingly didn't  do \cite{RovelliPartialObservables,DittrichPartialConstrained,TamborninoObservables}.  It would be interesting to explore relations between these ideas and the reformed notion of observables.


\section{First-Class Constraints and Gauge?}

It is generally accepted that first-class constraints are related to gauge freedom, but there are two main  views about what that precise relationship is.  The original view, which retains manifest equivalence to the Lagrangian and which disappeared as the 1950s wore on and started reappearing around 1980, is that gauge transformations are generated by a 
 \emph{tuned sum} of first-class  constraints (primary, secondary, \emph{etc.}), the ``gauge generator'' $G$   \cite{RosenfeldQG,AndersonBergmann,MukundaGaugeGenerator,CastellaniGaugeGenerator,ShepleyPonsSalisburyTurkish,SundermeyerSymmetries}.  
For Maxwell's electromagnetism the gauge generator is  $$ G(t)= \int d^3x (-\dot{\epsilon}(t,x) \pi^0 + \epsilon(t,x) \pi^i,_i(t,x)).$$
Expressions are also known for General Relativity, both without and with the $3+1$ split  \cite{AndersonBergmann,CastellaniGaugeGenerator} 
This $G$-based view competes with what became the majority  view (especially in books), that each first-class constraint  FC \emph{alone}  generates a gauge transformation  \cite{Bergmann,DiracLQM,Govaerts,HenneauxTeitelboim,RotheRothe}.
  
While gauge transformations are not this paper's primary concern, gauge transformations and observables are naturally interrelated:  observables ought to be invariant (or perhaps covariant) under gauge transformations, and transformations under which observables are invariant (or perhaps covariant) ought to be gauge transformations.  Thus a revision of the notion of gauge transformation calls for a revision of the definition of observables \cite{PonsSalisburySundermeyerFolklore}, and to some degree \emph{vice versa}.

Fortunately one  can \emph{test} definitions of observables  by calculation using two formulations of a theory, one without gauge freedom and one with gauge freedom.  The formulations, being empirically equivalent, must have equivalent observables.  The equivalence of non-gauge and gauge formulations of massive quantum electrodynamics is presupposed in quantum field theory to show that the theory is renormalizable (shown using the Stueckelberg-Utiyama gauge formulation with a gauge compensation field) and unitary (shown using in effect the de Broglie-Proca formulation)  \cite[pp. 738, 739]{PeskinSchroeder}\cite[chapter 21]{WeinbergQFT2}\cite[chapter 10]{Kaku}.  
 For de Broglie-Proca non-gauge massive electromagnetism, all constraints are second-class, so everything is observable.  Demanding equivalent observables from gauge Stueckelberg-Utiyama electromagnetism, one can ascertain whether observables should have $0$ Poisson bracket with each first-class constraint separately, or rather have $0$ Poisson bracket only with the gauge generator $G$.  It turns out that  the usual definition fails while the Pons-Salisbury-Sundermeyer definition with $G$ succeeds \cite{ObservablesEquivalentCQG}.   This result parallels arguments based on the requirement of Hamiltonian-Lagrangian equivalence \cite{FirstClassNotGaugeEM}.


\section{Internal \emph{vs.} External Gauge Symmetries and Invariance \emph{vs.} Covariance }
 
Definitions of observables often have been designed around  electromagnetism, an internal symmetry, and  imported into GR without much consideration for whether external symmetries might differ relevantly from internal ones  \cite{BergmannObservableNC,Bergmann,DiracLQM}.  In GR,  $G$, acting on the $4$-dimensional metric,  gives the $4$-d Lie derivative  \cite{CastellaniGaugeGenerator},  $$ \pounds_{\xi} g_{\mu\nu} = \xi^{\alpha} g_{\mu\nu},_{\alpha} + g_{\mu\alpha} \xi^{\alpha},_{\nu} + g_{\alpha\nu} \xi^{\alpha},_{\mu}.$$ The second and third terms are (for weak fields) analogous to the electromagnetic case, but the transport term $ \xi^{\alpha} g_{\mu\nu},_{\alpha}$ differentiates $g_{\mu\nu}$, thus making the transformation  ``external.''   

The use of $G$ does not suffice to yield changing and locally varying observables \cite{ObservablesEquivalentCQG}, at least not ones that one would expect on Lagrangian/geometric grounds such as the $4$-metric.  Because $G$ gives the $4$-d Lie derivative, the definition of observables $\{ O,  G[\xi^{\alpha}] \}=0$ $(\forall \xi^{\alpha})$ implies $\pounds_{\xi} O =0$  $(\forall \xi^{\alpha})$.  Observables are not allowed to change in any direction, so  $O$  is constant over time and space even with $G$. The problem of time is still present for observables even using $G$. At this point one might recall criticisms by Smolin and by Kucha\v{r} of the usual definition of observables, discussed above, as well as Bergmann's occasional view that observables should be local.    Can one devise a systematic definition of observables that can also encounter a crucial test with the right examples?

One might wish to amend Kucha\v{r}'s proposal in two ways (apart from using $G$ \cite{PonsSalisburySundermeyerFolklore}).  First, Kucha\v{r}'s common-sense argument against $\{ O, \mathcal{H}_0 \} =0$ is just as compelling against $\{ O, \mathcal{H}_i \} =0$ (which he retains), because spatial variation is as evident as change. Thus one should treat space and time alike and consider relaxing both $\{ O, \mathcal{H}_0 \} =0$ and   $\{ O, \mathcal{H}_i \} =0$.   Second, abolishing any restrictions at all on time gauge behavior is  unnecessarily strong, making  any arbitrary behavior regarding time gauge transformations  admissible.  There is an overlooked intermediate position, not invariance but covariance, imposing some well-defined time coordinate transformation rule (scalar, vector, \emph{etc.}).  Infinitesimally, one would thus expect (especially after embracing $G$) to have a $4$-dimensional Lie derivative, not $0$, be the result of the Poisson bracket in the definition of observables.  One also notices that whereas electromagnetic gauge transformations are ineffable mental acts with no operational correlate (no knob or reading on a voltmeter), so electromagnetic observables must be invariant, general relativistic gauge (coordinate) transformations are already familiar from geography and daylight savings time.  Being gauge-invariant does not require being the same at 1 a.m. Eastern Daylight Time and 1 a.m. Eastern Standard Time an hour later.   Hence covariance rather than invariance is a reasonable criterion \cite{GRChangeNoKilling}.


\section{Testing Definitions using Massive Gravity} 
 
As massive electromagnetism comes in  non-gauge and gauge versions, so does massive gravity.  Non-gauge versions appeared in nonlinear form in the 1960s  \cite{OP,FMS}.  Gauge versions can be achieved by parametrization, promoting (or perhaps demoting \cite{Kuchar73}) preferred coordinates into fields varied in the action principle  
\cite{Schmelzer,Arkani,MassiveGravity1}.  
Massive gravity was rejected in the early 1970s due to either instability (spin $2$-spin $0$) or a discontinuous massless limit \cite{vDVmass1,Zakharov,vDVmass2,DeserMass}.  Progress was made on both fronts during the 2000s \cite{Vainshtein2,deRhamGabadadze,HassanRosenNonlinear}, along with new challenges \cite{DeserWaldronAcausality}.  Fortunately, for present purposes it doesn't matter at all what problems massive gravity has.  What matters is the relationship between the non-gauge and gauge versions.  One might as well choose the version that makes the calculations the easiest, the Freund-Maheshwari-Schonberg (FMS) theory  \cite{FMS}, which, when parametrized, gives  nicest form, namely, \emph{minimal} coupling of the scalar clock fields \cite{Schmelzer}. 

As with de Broglie-Proca massive electromagnetism, the observables in  non-gauge massive gravity are obvious because all constraints are second-class \cite{PittsQG05}. Thus trivially everything has $0$ Poisson bracket with all first-class constraints,  making everything  observable, including the $4$-metric $g_{\mu\nu}$ and the  non-zero momenta $\pi^{mn}$.  The theory is  merely Poincar\'{e}-invariant. 
The FMS mass term is  $$\mathcal{L}_{m} = m^2  \sqrt{-g} + m^2 \sqrt{-\eta} - \frac{1}{2} m^2 \sqrt{-g} g^{\mu\nu} \eta_{\mu\nu},$$ where  $\eta_{\mu\nu}=diag(-1,1,1,1)$ in  Cartesian coordinates.  $\sqrt{-\eta}$ gives just a constant in the action.

One obtains the gauge version by parametrization, turning  preferred Cartesian coordinates into clock fields $X^A(x),$ functions of arbitrary coordinates $x^{\mu}.$  Only the mass term is affected.  
Now the reason for choosing the FMS massive theory becomes evident, namely, that its mass term, in contrast to the many other options out there (\emph{e.g.}, \cite{OP,HassanRosenNonlinear}), gives \emph{minimally coupled} scalar clock fields in  the expression  $$ \sqrt{-g} g^{\mu\nu} \eta_{AB} \frac{\partial X^A}{\partial x^{\mu}}  \frac{\partial X^B}{\partial x^{\nu}},$$ instead of using inverses, determinants, and/or fractional powers of $\eta_{AB} \frac{\partial X^A}{\partial x^{\mu}}  \frac{\partial X^B}{\partial x^{\nu}},$ or sums thereof. 
It turns out that one can do calculations in the general case anyway  \cite{KlusonHamiltonian}, but that is a pleasant surprise. 
The parametrized mass term is  $$\mathcal{L}_{mg} = m^2  \sqrt{-g} + m^2 \sqrt{-\eta} - \frac{1}{2} m^2 \sqrt{-g} g^{\mu\nu} \eta_{AB} \frac{\partial X^A}{\partial x^{\mu}}  \frac{\partial X^B}{\partial x^{\nu}},$$  $\eta_{AB}=diag(-1,1,1,1).$ (It isn't necessary to parametrize $\sqrt{-\eta}$ because the result is a total divergence.)  This is just a cosmological constant $\sqrt{-g},$ a harmless constant $\sqrt{-\eta},$ and four minimally coupled scalar fields (one with the wrong sign).  Gauge (parametrized) massive gravity becomes non-gauge massive gravity upon gauge-fixing $X^A -x^{\alpha}=0.$

Knowing that the $4$-metric (and hence the inverse metric $g^{\mu\nu}$) is observable in the non-gauge theory, one can \emph{demand} that observables in gauge massive gravity  be equivalent to the non-gauge massive gravity observables. The equivalent quantity is $g^{\mu\nu} X^A,_{\mu} X^B,_{\nu} :$ the gradients of the clock fields act as the tensor transformation law to the preferred Cartesian coordinates in which the non-gauge formulation is already expressed.  By seeing how the quantity $g^{\mu\nu} X^A,_{\mu} X^B,_{\nu} $ behaves in parametrized massive gravity, we can learn how observables behave under \emph{external} gauge transformations.  Should observables satisfy $\{ O, G[\xi] \}  = 0$, or $ \{ O, G[{\xi}] \} \sim  \pounds_{\xi}O $ (possibly just on-shell)?  


\subsection{Hamiltonian for Gauge Massive Gravity}  

For General Relativity with  minimally coupled scalar fields  and a cosmological constant, the Poisson bracket `algebra' of constraints is just  as in GR \cite{Sundermeyer}.  For the parametrized version of the Freund-Maheshwarei-Schonberg theory, the same result therefore holds.  It is straightforward to take the parametrized Lagrangian density and perform the constrained  Legendre transformation with 4 minimally coupled scalars and $\Lambda$   GR \cite{MTW,Sundermeyer,Wald}.  In the ADM $3+1$ split, the $4$-metric is broken into the  lapse $N$, the shift vector $\beta^{i},$ and spatial metric $h_{ij}.$  There are new canonical momenta for the clock fields:  $$ \pi_A = \frac{\partial \mathcal{L}_{mg} }{\partial X^A,_0 } =  -m^2 \mathfrak{g}^{\mu0} \eta_{AB} X^B,_{\mu}.$$  Inverting, one gets  $$ \dot{X}^A = N  \pi_B \eta^{AB} m^{-2}/ \sqrt{h}  + \beta^i X^A,_i.$$ 
  The Hamiltonian density is 
{ \begin{eqnarray*} \mathcal{H}_{mg} = N  \left(\mathcal{H}_0    -m^2 \sqrt{h} + \frac{ \pi_A \pi_B \eta^{AB} }{2 m^2 \sqrt{h} } +  \frac{m^2}{2} \sqrt{h} h^{ij} X^A,_i X^B,_j \eta_{AB} \right) \\ + \beta^i (\mathcal{H}_i+  X^A,_i \pi_A) - m^2 \sqrt{-\eta}, \end{eqnarray*} \\ with   $\mathcal{H}_0$ and $\mathcal{H}_i$ as in GR.  }  This expression has the same form as in GR (apart from an irrelevant constant $\sqrt{-\eta}$) if one defines a total (gravitational plus matter) Hamiltonian constraint $$\mathcal{H}_{0T} = \mathcal{H}_0    -m^2 \sqrt{h} + \frac{ \pi_A \pi_B \eta^{AB} }{2 m^2 \sqrt{h} } +  \frac{m^2}{2} \sqrt{h} h^{ij} X^A,_i X^B,_j \eta_{AB} $$ and a total momentum constraint $$\mathcal{H}_{iT}= \mathcal{H}_i+  X^A,_i \pi_A.$$ The Hamiltonian for parametrized massive gravity is (apart from terms involving primary constraints)   $$ \mathcal{H}_{mg} = N \mathcal{H}_{0T}  + \beta^i \mathcal{H}_{iT} -m^2 \sqrt{-\eta}.$$


\subsection{Applying the Gauge Generator in Massive Gravity}

  Avoiding velocities  requires $3+1$ split of coordinate transformation descriptor $\xi^{\mu}$  \cite{CastellaniGaugeGenerator,PonsSalisburyShepleyYang}:   $\epsilon = N \xi^0$ is primitive and so has $0$ Poisson brackets; the same holds for  $\epsilon^i = \xi^i + \beta^i \xi^0.$  The primary constraints are as in General Relativity:  $p$  conjugate to $N$ and $p_i$  conjugate to $\beta^i$ both vanish.   
The generator of changes of time coordinate in vacuum General Relativity is $$G[\epsilon, \dot{\epsilon}] = \int d^3x [\epsilon \mathcal{H}_0  + \epsilon p_j h^{ij} N,_i + \epsilon(N p_i h^{ij}),_j + \epsilon (p N^j),_j + \dot{\epsilon} p].$$
This entity generates on phase space $\times$ time a transformation that, for solutions of the Hamiltonian field equations, changes the time coordinate in accord with  $4$-dimensional tensors.  Given how the gauge generator can be built algorithmically starting with the primary constraints \cite{PonsDirac}, one would expect the same expression for the gauge generator for parametrized massive FMS gravity, only with matter included in the secondary constraints.  The Hamiltonian takes the form of  GR + $\Lambda$ + minimally coupled scalars with altered matter-containing constraints $\mathcal{H}_{0T}$ and $\mathcal{H}_{iT}.$  

One can verify that the resulting modified expression for $G_T$ indeed generates gauge transformations; indeed displaying that calculation in more detail than appeared previously \cite{ObservablesEquivalentCQG} is the aim of this paper. 
 For the space-time metric there is no difference because matter does not couple to gravitational momenta.  For the new matter fields one has 
$$ \{ G[\epsilon, \dot{\epsilon}], X^A(y) \} = - \epsilon(y) \pi_B \eta^{BA} /(m^2 \sqrt{h}) =  -\xi^0 N \pi_B \eta^{BA} /(m^2 \sqrt{h}).$$ 
Using the relation $\dot{X}^A = N  \pi_C \eta^{AC} m^{-2}/ \sqrt{h}  + \beta^i X^A,_i$ recovered from $\dot{X}^A= \frac{\delta H}{\delta \pi_A},$ one gets
 $$\{ G[\epsilon, \dot{\epsilon}], X^A(y) \} = - \xi^0 X^A,_0 + \xi^0 \beta^i X^A,_i $$ {on-shell}. This relates nicely to  the Lie derivative of the scalar clock fields.  The second term is involved in a cancellation.   

%

  
 The spatial gauge generator for vacuum GR is \cite{CastellaniGaugeGenerator} $$ G[\epsilon^i, \dot{\epsilon}^i] = \int d^3x [\epsilon^i \mathcal{H}_i  +  \epsilon^i N^j,_i p_j - \epsilon^j,_i N^i p_j  + \epsilon^i N,_i p + \epsilon^i,_0 p_i].$$  It generates $3$-d spatial  Lie derivatives of the $4$-metric $g_{\mu\nu}$ even off-shell.  Making the obvious alteration of the secondary constraint to include matter through  $\mathcal{H}_{iT}$  gives the correct gauge generator, giving a Lie derivative of the scalar clock fields:  
  $$\{ G_T[\epsilon^i, \dot{\epsilon}^i], X^A(y) \} = \{ \int d^3x \epsilon^i(x) X^C,_i \pi_C,  X^A(y) \}.$$  Going on-shell one gets the result $$  - (\xi^i + \beta^i \xi^0) X^A,_i.$$ The  off-shell result will be worked out below.

The full gauge generator $G_T$ is the sum of these two parts \cite{CastellaniGaugeGenerator}:  $$  G_T[\epsilon, \dot{\epsilon}] + G_T[\epsilon^i, \dot{\epsilon}^i]$$ 
The vacuum gauge generator combination generates $4$-dimensional coordinate transformations on the space-time metric, at least for solutions: $$ \{ G[\epsilon, \dot{\epsilon}] + G[\epsilon^i, \dot{\epsilon}^i], g^{\mu\nu} \} = - \pounds_{\xi} g^{\mu\nu} $$ \cite{CastellaniGaugeGenerator,PonsSalisburyShepleyYang} (on-shell) in General Relativity.  The new material parts of  the total momentum constraint and total Hamiltonian constraint have no gravitational momenta and hence do not affect the space-time metric.

 Acting on the clock fields $X^A,$ the total generator  gives (going on-shell eventually) 
\begin{eqnarray*} 
 \{ G_T[\epsilon, \dot{\epsilon}] + G_T[\epsilon^i, \dot{\epsilon}^i], X^A(y) \} =  - \xi^0 X^A,_0 + \xi^0 \beta^i X^A,_i - (\xi^i + \beta^i \xi^0) X^A,_i  \\ = - \xi^0 X^A,_0  - \xi^i X^A,_i = - \xi^{\mu} X^A,_{\mu}, \end{eqnarray*} 
the proper $4$-dimensional expression for (minus) the Lie derivative of a scalar field with respect to the space-time vector field $\xi^{\mu}$ describing the infinitesimal coordinate transformation.  One can thus see in outline how the whole of $g^{\mu\nu} X^A,_{\mu} X^B,_{\nu}$ behaves nicely, at least on-shell.


\subsection{Off-Shell Calculation in Detail}

  It now being clear in general outline what to expect, one can profitably do the calculation with more detail and while remaining off-shell. 
 The matter-inclusive  spatial gauge generator is $$ G_T[\epsilon^i, \dot{\epsilon}^i] = \int d^3x [\epsilon^i \mathcal{H}_{iT}  +  \epsilon^i N^j,_i p_j - \epsilon^j,_i N^i p_j  + \epsilon^i N,_i p + \epsilon^i,_0 p_i].$$  One therefore has
\begin{eqnarray} \{G_T[\epsilon^i, \dot{\epsilon}^i], g^{\mu\nu} X^A,_{\mu} X^B,_{\nu}\} =  \{G_T[\epsilon^i, \dot{\epsilon}^i], g^{\mu\nu} \}  X^A,_{\mu} X^B,_{\nu}    + g^{\mu\nu}  \{G_T[\epsilon^i, \dot{\epsilon}^i], X^A,_{\mu} \} X^B,_{\nu} +  B\leftrightarrow A  \nonumber \end{eqnarray}  using the Leibniz product rule, 
\begin{eqnarray} 
=   \{G_T[\epsilon^i, \dot{\epsilon}^i], g^{\mu\nu} \}  X^A,_{\mu} X^B,_{\nu}    + (g^{m\nu}  \{G_T[\epsilon^i, \dot{\epsilon}^i], X^A,_{m} \} + g^{0\nu}  \{G_T[\epsilon^i, \dot{\epsilon}^i], X^A,_{0} \})  X^B,_{\nu}      
+ B \leftrightarrow A    \nonumber  \end{eqnarray} 
 splitting up space-time into space and time,
\begin{eqnarray} 
=  \{G_T[\epsilon^i, \dot{\epsilon}^i], g^{\mu\nu} \}  X^A,_{\mu} X^B,_{\nu}    + (g^{m\nu}  \frac{\partial}{\partial x^m} \{G_T[\epsilon^i, \dot{\epsilon}^i], X^A \} + g^{0\nu}  \frac{\partial}{\partial x^0} \{G_T[\epsilon^i, \dot{\epsilon}^i], X^A \})  X^B,_{\nu}     
+ B \leftrightarrow A  \nonumber
\end{eqnarray}  pulling out the spatial derivative \cite[p. 58]{ThiemannLectures} and  using the Anderson-Bergmann velocity Poisson bracket for the 0th component \cite{AndersonBergmann},  
%
\begin{eqnarray} 
=   \{G_T[\epsilon^i, \dot{\epsilon}^i], g^{\mu\nu} \}  X^A,_{\mu} X^B,_{\nu}    + [g^{m\nu}  \frac{\partial}{\partial x^m}   (-\epsilon^i  X^A,_i )
+ g^{0\nu}  \frac{\partial}{\partial x^0}  (-\epsilon^i  X^A,_i )   ]  X^B,_{\nu}      
+ B \leftrightarrow A   \nonumber  \end{eqnarray} 
using the explicit form of the matter-enriched spatial gauge generator,   
\begin{eqnarray} 
= \{G_T[\epsilon^i, \dot{\epsilon}^i], g^{\mu\nu} \}  X^A,_{\mu} X^B,_{\nu}    + [- g^{m\nu}  \pounds_{\epsilon}  X^A,_m
+ g^{0\nu}  \frac{\partial}{\partial x^0}  (-\pounds_{\epsilon}  X^A ) ]  X^B,_{\nu}      
+ B \leftrightarrow A  \nonumber \end{eqnarray}
using the commutation of Lie and partial derivatives  \cite[p. 105]{Schouten} as applied to space rather than space-time.

The generator of time coordinate transformations is $$G_T[\epsilon, \dot{\epsilon}] = \int d^3x [\epsilon \mathcal{H}_{0T}  + \epsilon p_j h^{ij} N,_i + \epsilon(N p_i h^{ij}),_j + \epsilon (p N^j),_j + \dot{\epsilon} p].$$ 
The temporal gauge generator thus acts on  $g^{\mu\nu} X^A,_{\mu} X^B,_{\nu}$ as 
\begin{eqnarray} \{G_T[\epsilon, \dot{\epsilon}], g^{\mu\nu} X^A,_{\mu} X^B,_{\nu}\} =
\{G_T[\epsilon, \dot{\epsilon}], g^{\mu\nu}\}  X^A,_{\mu} X^B,_{\nu} +  \{G_T[\epsilon, \dot{\epsilon}],   X^A,_{\mu}  \} X^B,_{\nu} g^{\mu\nu} +   B \leftrightarrow A  \nonumber  
\end{eqnarray} 
by the Leibniz product rule,
\begin{eqnarray} 
= \{G_T[\epsilon, \dot{\epsilon}], g^{\mu\nu}\}  X^A,_{\mu} X^B,_{\nu} +  \{G_T[\epsilon, \dot{\epsilon}],   X^A \},_{\mu}  X^B,_{\nu} g^{\mu\nu} +   B \leftrightarrow A   \nonumber  
\end{eqnarray} 
by pulling out the spatial derivative and using the Anderson-Bergmann velocity Poisson bracket.
Using the result for the clock fields $ \{ G_T[\epsilon, \dot{\epsilon}], X^A(y) \} = - \epsilon(y) \pi_B \eta^{BA} /(m^2 \sqrt{h}),$ 
one infers
\begin{eqnarray} \{G_T[\epsilon, \dot{\epsilon}], g^{\mu\nu} X^A,_{\mu} X^B,_{\nu}\}
 =
\{G_T[\epsilon, \dot{\epsilon}], g^{\mu\nu}\}  X^A,_{\mu} X^B,_{\nu} +  (   -\epsilon \pi^A  /(m^2 \sqrt{h})  ),_{\mu}  X^B,_{\nu} g^{\mu\nu} +   B \leftrightarrow A.    \nonumber 
\end{eqnarray}

The combined spatio-temporal gauge generator thus yields
\begin{eqnarray} 
=   \{G_T[\epsilon^i, \dot{\epsilon}^i] + G_T[\epsilon, \dot{\epsilon}], g^{\mu\nu}   X^A,_{\mu} X^B,_{\nu} \} = -(\pounds_{\xi} g^{\mu\nu}) X^A,_{\mu} X^B,_{\nu} 
- g^{\mu\nu} X^B,_{\nu} \frac{\partial}{\partial x^{\mu} } \pounds_{\epsilon}X^A  \nonumber \\
- B \leftrightarrow A 
- (\epsilon \pi^A m^{-2} \sqrt{h} ),_{\mu} X^B,_{\nu} g^{\mu\nu} - B \leftrightarrow A  \nonumber   \\
=
-(\pounds_{\xi} g^{\mu\nu}) X^A,_{\mu} X^B,_{\nu} 
- g^{\mu\nu} X^B,_{\nu} \frac{\partial}{\partial x^{\mu} } ( \xi^i X^A,_i+ \beta^i \xi^0 X^A,_i  + N \xi^0  \pi^A m^{-2} \sqrt{h} ) - B \leftrightarrow A  \nonumber \\  %
=
-(\pounds_{\xi} g^{\mu\nu}) X^A,_{\mu} X^B,_{\nu} 
- g^{\mu\nu} X^B,_{\nu} \frac{\partial}{\partial x^{\mu} } \left( \xi^i X^A,_i + \xi^0  \frac{\delta H}{\delta \pi_A} \right) - B \leftrightarrow A  \nonumber   \\
=
-(\pounds_{\xi} g^{\mu\nu}) X^A,_{\mu} X^B,_{\nu} 
- g^{\mu\nu} X^B,_{\nu} \frac{\partial}{\partial x^{\mu} } \left( \xi^{\nu} X^A,_{\nu}  - \xi^0 X^A,_0 + \xi^0 \frac{\delta H}{\delta \pi_A} \right) - B \leftrightarrow A  \nonumber   \\
=
-(\pounds_{\xi} g^{\mu\nu} X^A,_{\mu} X^B,_{\nu}) 
- g^{\mu\nu} X^B,_{\nu} \frac{\partial}{\partial x^{\mu} } \left(   - \xi^0 X^A,_0 + \xi^0 \frac{\delta H}{\delta \pi_A} \right) - B \leftrightarrow A  \nonumber   \end{eqnarray}
The term that is not a $4$-dimensional Lie derivative vanishes on-shell.

By the equivalence of the non-gauge and gauge observables, $g^{\mu\nu} X^A,_{\mu} X^B,_{\nu}$ must be observable in the gauge theory because $g^{\mu\nu}$ is observable in the non-gauge theory.  Knowing that $g^{\mu\nu} X^A,_{\mu} X^B,_{\nu}$ must be an observable and calculating how $g^{\mu\nu} X^A,_{\mu} X^B,_{\nu}$ is acted upon by $G$, we learn  that observables  give a Lie derivative rather than $0$ when one takes their Poisson bracket with $G$.  
 Thus for observables one has  $$\{ G, O \} = -\pounds_{\xi} O \neq 0 $$ on-shell, when $G$ generates coordinate transformations, an external symmetry.

 Covariance, not invariance, suffices for the external gauge symmetry in this case.  Thus the usual vanishing $0$ Poisson bracket condition is wrong at least in this case.  This case looks just like GR plus $\Lambda$ plus minimally coupled scalars, so the same result should hold there.  Adding   $\Lambda$ and  minimally coupled scalars to GR is insignificant, so the same result should hold for GR itself.  Thus quantities that change by a Lie derivative under Poisson bracket with the gauge generator are observable:  scalar fields, vector fields, tensors, densities, \emph{etc.}



\begin{thebibliography}{}

\bibitem[Anderson, 1962]{AndersonChange}
Anderson, J.~L. (1962).
\newblock Absolute change in general relativity.
\newblock In {\em Recent Developments in General Relativity}, pages 121--126.
  Pergamon and PWN, Oxford and Warsaw.

\bibitem[Anderson and Bergmann, 1951]{AndersonBergmann}
Anderson, J.~L. and Bergmann, P.~G. (1951).
\newblock Constraints in covariant field theories.
\newblock {\em Physical Review}, 83:1018--1025.

\bibitem[Arkani-Hamed et~al., 2003]{Arkani}
Arkani-Hamed, N., Georgi, H., and Schwartz, M.~D. (2003).
\newblock Effective field theory for massive gravitons and gravity in theory
  space.
\newblock {\em Annals of Physics}, 305:96--118.
\newblock hep-th/0210184v2.

\bibitem[Bergmann, 1956]{BergmannObservableNC}
Bergmann, P.~G. (1956).
\newblock Introduction of ``true observables'' into the quantum field
  equations.
\newblock {\em Il Nuovo Cimento}, 3:1177--1185.

\bibitem[Bergmann, 1961]{Bergmann}
Bergmann, P.~G. (1961).
\newblock Observables in general relativity.
\newblock {\em Reviews of Modern Physics}, 33:510--514.

\bibitem[Bergmann, 1962]{BergmannHandbuch}
Bergmann, P.~G. (1962).
\newblock The general theory of relativity.
\newblock In Fl\"{u}gge, S., editor, {\em Prinzipien der Elektrodynamik und
  Relativit\"{a}tstheorie}, volume~IV of {\em Handbuch der Physik}, pages
  203--272. Springer, Berlin.

\bibitem[Bergmann and Goldberg, 1955]{BergmannGoldberg}
Bergmann, P.~G. and Goldberg, I. (1955).
\newblock Dirac bracket transformations in phase space.
\newblock {\em Physical Review}, 98:531--538.

\bibitem[Bergmann and Komar, 1962]{BergmannKomarRoyaumont}
Bergmann, P.~G. and Komar, A. (1962).
\newblock Observables and commutation relations.
\newblock In {\em Les Th\'{e}ories Relativistes de la Gravitation, Royaumont,
  21-27 Juin 1959}, pages 309--325. Centre National de la Recherche
  Scientifique, Paris.

\bibitem[Bergmann and Schiller, 1953]{BergmannSchiller}
Bergmann, P.~G. and Schiller, R. (1953).
\newblock Classical and quantum field theories in the {Lagrangian} formalism.
\newblock {\em Physical Review}, 89:4--16.

\bibitem[Boulware and Deser, 1972]{DeserMass}
Boulware, D.~G. and Deser, S. (1972).
\newblock Can gravitation have a finite range?
\newblock {\em Physical Review D}, 6:3368--3382.

\bibitem[Castellani, 1982]{CastellaniGaugeGenerator}
Castellani, L. (1982).
\newblock Symmetries in constrained {Hamiltonian} systems.
\newblock {\em Annals of Physics}, 143:357--371.

\bibitem[{de Rham} et~al., 2011]{deRhamGabadadze}
{de Rham}, C., Gabadadze, G., and Tolley, A.~J. (2011).
\newblock Resummation of massive gravity.
\newblock {\em Physical Review Letters}, 106:231101.
\newblock arXiv:1011.1232v2 [hep-th].

\bibitem[Deffayet et~al., 2002]{Vainshtein2}
Deffayet, C., Dvali, G., Gabadadze, G., and Vainshtein, A.~I. (2002).
\newblock Nonperturbative continuity in graviton mass versus perturbative
  discontinuity.
\newblock {\em Physical Review D}, 65:044026.
\newblock hep-th/0106001v2.

\bibitem[Deser and Waldron, 2013]{DeserWaldronAcausality}
Deser, S. and Waldron, A. (2013).
\newblock Acausality of massive gravity.
\newblock {\em Physical Review Letters}, 110:111101.
\newblock arXiv:1212.5835.

\bibitem[Dirac, 1964]{DiracLQM}
Dirac, P. A.~M. (1964).
\newblock {\em Lectures on Quantum Mechanics}.
\newblock Belfer Graduate School of Science, Yeshiva University.
\newblock Dover reprint, Mineola, New York, 2001.

\bibitem[Dittrich, 2007]{DittrichPartialConstrained}
Dittrich, B. (2007).
\newblock Partial and complete observables for {Hamiltonian} constrained
  systems.
\newblock {\em General Relativity and Gravitation}, 39:1891--1927.
\newblock arXiv:gr-qc/0411013.

\bibitem[Freund et~al., 1969]{FMS}
Freund, P. G.~O., Maheshwari, A., and Schonberg, E. (1969).
\newblock Finite-range gravitation.
\newblock {\em Astrophysical Journal}, 157:857--867.

\bibitem[Govaerts, 1991]{Govaerts}
Govaerts, J. (1991).
\newblock {\em Hamiltonian Quantisation and Constrained Dynamics}.
\newblock Leuven Notes in Mathematical and Theoretical Physics 4B. Leuven
  University Press, Leuven.

\bibitem[Hassan and Rosen, 2011]{HassanRosenNonlinear}
Hassan, S.~F. and Rosen, R.~A. (2011).
\newblock On non-linear actions for massive gravity.
\newblock {\em Journal of High Energy Physics}, 1107(009).
\newblock arXiv:1103.6055v3 [hep-th].

\bibitem[Henneaux and Teitelboim, 1992]{HenneauxTeitelboim}
Henneaux, M. and Teitelboim, C. (1992).
\newblock {\em Quantization of Gauge Systems}.
\newblock Princeton University Press, Princeton.

\bibitem[Isham, 1993]{IshamTime}
Isham, C.~J. (1993).
\newblock Canonical quantum gravity and the problem of time.
\newblock In Ibort, L.~A. and Rodr\'{i}guez, M.~A., editors, {\em Integrable
  Systems, Quantum Groups, and Quantum Field Theories}, pages 157--287. Kluwer,
  Dordrecht.
\newblock {Lectures presented at the NATO Advanced Study Institute ``Recent
  Problems in Mathematical Physics,'' Salamanca, June 15-27, 1992;
  gr-qc/9210011}.

\bibitem[Kaku, 1993]{Kaku}
Kaku, M. (1993).
\newblock {\em Quantum Field Theory: A Modern Introduction}.
\newblock Oxford University, New York.

\bibitem[Kiefer, 2012]{Kiefer3rd}
Kiefer, C. (2012).
\newblock {\em Quantum Gravity}.
\newblock Oxford University Press, 3rd edition.

\bibitem[Kluso\v{n}, 2011]{KlusonHamiltonian}
Kluso\v{n}, J. (2011).
\newblock Hamiltonian analysis of the {Higgs} mechanism for graviton.
\newblock {\em Classical and Quantum Gravity}, 28:155014.
\newblock arXiv:1005.5458 [hep-th].

\bibitem[Kucha\v{r}, 1973]{Kuchar73}
Kucha\v{r}, K. (1973).
\newblock Canonical quantization of gravity.
\newblock In Israel, W., editor, {\em Relativity, Astrophysics, and Cosmology},
  pages 237--288. D. Reidel, Dordrecht.

\bibitem[Kucha\v{r}, 1992]{KucharCanadian92}
Kucha\v{r}, K.~V. (1992).
\newblock Time and interpretations of quantum gravity.
\newblock In Kunstatter, G., Vincent, D., and Williams, J., editors, {\em
  Proceedings of the 4th Canadian Conference on General Relativity and
  Relativistic Astrophysics}, pages 211--314. World Scientific, Singapore.

\bibitem[Kucha\v{r}, 1993]{KucharCanonical93}
Kucha\v{r}, K.~V. (1993).
\newblock Canonical quantum gravity.
\newblock In Gleiser, R.~J., Kozameh, C.~N., and Moreschi, O.~M., editors, {\em
  General Relativity and Gravitation 1992: Proceedings of the Thirteenth
  International Conference on General Relativity and Gravitation held at
  Cordoba, Argentina, {28 June--4 July} 1992}, pages 119--150. Institute of
  Physics Publishing, Bristol.
\newblock arXiv:gr-qc/9304012.

\bibitem[Misner et~al., 1973]{MTW}
Misner, C., Thorne, K., and Wheeler, J.~A. (1973).
\newblock {\em Gravitation}.
\newblock Freeman, New York.

\bibitem[Mukunda, 1980]{MukundaGaugeGenerator}
Mukunda, N. (1980).
\newblock Generators of symmetry transformations for constrained {Hamiltonian}
  systems.
\newblock {\em Physica Scripta}, 21:783--791.

\bibitem[Ogievetsky and Polubarinov, 1965]{OP}
Ogievetsky, V.~I. and Polubarinov, I.~V. (1965).
\newblock Interacting field of spin 2 and the {Einstein} equations.
\newblock {\em Annals of Physics}, 35:167--208.

\bibitem[Peskin and Schroeder, 1995]{PeskinSchroeder}
Peskin, M.~E. and Schroeder, D.~V. (1995).
\newblock {\em An Introduction to Quantum Field Theory}.
\newblock Addison-Wesley, Reading, Massachusetts.

\bibitem[Pitts, 2006]{PittsQG05}
Pitts, J.~B. (2006).
\newblock Constrained dynamics of universally coupled massive spin 2-spin 0
  gravities.
\newblock {\em Journal of Physics: Conference Series}, 33:279--284.
\newblock hep-th/0601185, Talk given at QG05, Cala Gonone, Sardinia, Italy,
  September 2005.

\bibitem[Pitts, 2014a]{GRChangeNoKilling}
Pitts, J.~B. (2014a).
\newblock Change in {Hamiltonian} general relativity from the lack of a
  time-like {Killing} vector field.
\newblock {\em Studies in History and Philosophy of Modern Physics}, 47:68--89.
\newblock arXiv:1406.2665.

\bibitem[Pitts, 2014b]{FirstClassNotGaugeEM}
Pitts, J.~B. (2014b).
\newblock A first class constraint generates not a gauge transformation, but a
  bad physical change: {The} case of electromagnetism.
\newblock {\em Annals of Physics}, 351:382--406.
\newblock arXiv:1310.2756.

\bibitem[Pitts, 2017]{ObservablesEquivalentCQG}
Pitts, J.~B. (2017).
\newblock Equivalent theories redefine {Hamiltonian} observables to exhibit
  change in {General Relativity}.
\newblock {\em Classical and Quantum Gravity}, 34(055008).
\newblock doi.org/10.1088/1361-6382/aa5ce8; arXiv:1609.04812 [gr-qc].

\bibitem[Pitts and Schieve, 2007]{MassiveGravity1}
Pitts, J.~B. and Schieve, W.~C. (2007).
\newblock Universally coupled massive gravity.
\newblock {\em Theoretical and Mathematical Physics}, 151:700--717.
\newblock arXiv:gr-qc/0503051v3.

\bibitem[Pons, 2005]{PonsDirac}
Pons, J.~M. (2005).
\newblock On {Dirac's} incomplete analysis of gauge transformations.
\newblock {\em Studies in History and Philosophy of Modern Physics},
  36:491--518.
\newblock arXiv:physics/0409076v2.

\bibitem[Pons et~al., 2000]{PonsSalisburyShepleyYang}
Pons, J.~M., Salisbury, D.~C., and Shepley, L.~C. (2000).
\newblock Gauge transformations in {Einstein-Yang-Mills} theories.
\newblock {\em Journal of Mathematical Physics}, 41:5557--5571.
\newblock arXiv:gr-qc/9912086.

\bibitem[Pons et~al., 2010]{PonsSalisburySundermeyerFolklore}
Pons, J.~M., Salisbury, D.~C., and Sundermeyer, K.~A. (2010).
\newblock Observables in classical canonical gravity: {Folklore} demystified.
\newblock {\em Journal of Physics: Conference Series}, 222:012018.
\newblock First Mediterranean Conference on Classical and Quantum Gravity
  (MCCQG 2009); arXiv:1001.2726v2 [gr-qc].

\bibitem[Rosenfeld, 1930]{RosenfeldQG}
Rosenfeld, L. (1930).
\newblock {Zur Quantelung der Wellenfelder}.
\newblock {\em Annalen der Physik}, 397:113--152.
\newblock Translation by {Donald Salisbury and Kurt Sundermeyer}, ``On the
  Quantization of Wave Fields,'' \emph{European Physical Journal H} {\bf 42}
  (2017), pp. 63-94, doi:10.1140/epjh/e2016-70041-3.

\bibitem[Rothe and Rothe, 2010]{RotheRothe}
Rothe, H.~J. and Rothe, K.~D. (2010).
\newblock {\em Classical and Quantum Dynamics of Constrained Hamiltonian
  Systems}.
\newblock World Scientific, Hackensack, New Jersey.

\bibitem[Rovelli, 2002]{RovelliPartialObservables}
Rovelli, C. (2002).
\newblock Partial observables.
\newblock {\em Physical Review D}, 65:124013.
\newblock arXiv:gr-qc/0110035.

\bibitem[Schmelzer, 2000]{Schmelzer}
Schmelzer, I. (2000).
\newblock General ether theory.
\newblock {\em www.arxiv.org}, arXiv:gr-qc/0001101.

\bibitem[Schouten, 1954]{Schouten}
Schouten, J.~A. (1954).
\newblock {\em Ricci-Calculus: An Introduction to Tensor Analysis and Its
  Geometrical Applications}.
\newblock Springer, Berlin, second edition.
\newblock http://link.springer.com/book/10.1007

\bibitem[Shepley et~al., 2000]{ShepleyPonsSalisburyTurkish}
Shepley, L.~C., Pons, J.~M., and Salisbury, D.~C. (2000).
\newblock Gauge transformations in general relativity---{A} report.
\newblock {\em Turkish Journal of Physics}, 24(3):445--452.
\newblock Regional Conference on Mathematical Physics IX, 9-14 Aug. 1999,
  Istanbul, Turkey.

\bibitem[Smolin, 2001]{SmolinPresent}
Smolin, L. (2001).
\newblock The present moment in quantum cosmology: {Challenges} to the
  arguments for the elimination of time.
\newblock {\em www.arxiv.org, gr-qc/0104097}.
\newblock Slightly revised version of essay published in {Robin Durie,
  \emph{ed.}, \emph{Time and the Instant}, Clinamen Press, Manchester (2000)
  pp. 112-143}.

\bibitem[Sundermeyer, 1982]{Sundermeyer}
Sundermeyer, K. (1982).
\newblock {\em Constrained Dynamics: With Applications to Yang--Mills Theory,
  General Relativity, Classical Spin, Dual String Model}.
\newblock Springer, Berlin.
\newblock Lecture Notes in Physics, volume 169.

\bibitem[Sundermeyer, 2014]{SundermeyerSymmetries}
Sundermeyer, K. (2014).
\newblock {\em Symmetries in Fundamental Physics}.
\newblock Springer, Heidelberg, second edition.

\bibitem[Tambornino, 2012]{TamborninoObservables}
Tambornino, J. (2012).
\newblock Relational observables in gravity: {A} review.
\newblock {\em SIGMA}, 8(017).
\newblock arXiv:1109.0740.

\bibitem[Thiemann, 2003]{ThiemannLectures}
Thiemann, T. (2003).
\newblock Lectures on loop quantum gravity.
\newblock In Giulini, D. J.~W., Kiefer, C., and L\"{a}mmerzahl, C., editors,
  {\em Quantum Gravity: From Theory to Experimental Search}, number 631 in
  Lecture Notes in Physics, pages 41--135. Springer, Berlin.
\newblock gr-qc/0210094, based on lectures given at the 271st WE Heraeus
  Seminar ``Aspects of Quantum Gravity: From Theory to Experimental Search'',
  Bad Honnef, Germany, February 25th -- March 1st, 2002.

\bibitem[Torre, 1993]{TorreObservable}
Torre, C.~G. (1993).
\newblock Gravitational observables and local symmetries.
\newblock {\em Physical Review D}, 48:R2373--R2376.

\bibitem[van Dam and Veltman, 1970]{vDVmass1}
van Dam, H. and Veltman, M. (1970).
\newblock Massive and mass-less {Yang-Mills} and gravitational fields.
\newblock {\em Nuclear Physics B}, 22:397--411.

\bibitem[van Dam and Veltman, 1972]{vDVmass2}
van Dam, H. and Veltman, M. (1972).
\newblock On the mass of the graviton.
\newblock {\em General Relativity and Gravitation}, 3:215--220.

\bibitem[Wald, 1984]{Wald}
Wald, R.~M. (1984).
\newblock {\em General Relativity}.
\newblock University of Chicago, Chicago.

\bibitem[Weinberg, 1996]{WeinbergQFT2}
Weinberg, S. (1996).
\newblock {\em The Quantum Theory of Fields, Volume II: Modern Applications}.
\newblock Cambridge University Press, Cambridge.

\bibitem[Zakharov, 1970]{Zakharov}
Zakharov, V.~I. (1970).
\newblock Linearized gravitation theory and the graviton mass.
\newblock {\em Journal of Experimental and Theoretical Physics Letters},
  12:312--314.

\end{thebibliography}
\end{document}